\documentclass[journal]{IEEEtran}

\usepackage{amsmath,amssymb,amsfonts}
\usepackage{algorithmic}
\usepackage{graphicx}
\usepackage{textcomp}
\usepackage{booktabs}
\usepackage{array}
\usepackage{xcolor}
\usepackage{url}
\usepackage{tikz}
\usetikzlibrary{arrows.meta,positioning,calc,decorations.pathmorphing}
\usepackage[hidelinks]{hyperref}

\tikzset{
  joint/.style={circle,draw,fill=white,minimum size=4pt,inner sep=0pt},
  link/.style={line width=2.2pt,line cap=round,gray!75},
  sig/.style={-{Latex[length=2mm]},line width=0.6pt}
}

\newcommand{\unit}[1]{\,\mathrm{#1}}
\newcommand{\dd}{\hat{d}}
\newcommand{\Ieff}{I_\text{eff}}
\newcommand{\beff}{b_\text{eff}}
\newcommand{\taumax}{\tau_\text{max}}
\newcommand{\sigeff}{\sigma_\text{eff}}

\begin{document}

\title{Offset-Free Patient-Torque Estimation for a Knee\\
Rehabilitation Exoskeleton, and Two Failure Modes\\
of Its Predictive Safety Layer}

\author{Yongyan~Cao,~Jinshan~Tang~and~Xiaobo~Li%
\thanks{Y.~Cao and X.~Li are with Voryx Robotics, and J.~Tang is with George Mason University, Dept. of Health Administration and Policy. Corresponding author: Y.~Cao (e-mail: yongyancao@gmail.com).}}

\maketitle

\begin{abstract}
Safe rehabilitation is an interaction-dynamics problem: the controller must regulate a prescribed motion while absorbing involuntary spasm, voluntary effort, actuator compliance, and model mismatch as disturbances. This paper instantiates a predictive interaction-dynamics formulation on a series-elastic-actuated (SEA) knee joint, and then audits which of its layers the evidence actually supports. SEA feedforward reduces the gravity-compensated knee to a scalar double integrator; a noisy dynamic residual from spring deflection and finite-differenced acceleration, computed with deliberately mismatched inertia and damping, supplies an interaction-disturbance observation; a steady-state target converts the estimate into a cancelling input; and a finite-horizon quadratic program regulates the remaining deviation under range-of-motion, torque, and velocity constraints. Under a motion-opposing $15\unit{Nm}$ step with stiffness and damping matched across controllers, the estimator and target reduce steady-state error from $499.59$ to $0.70\unit{mrad}$, against $3.26\unit{mrad}$ for a best-tuned plain integral-action law --- whose full-run RMS, however, never falls below $157\unit{mrad}$ at any gain, against $3.09\unit{mrad}$ here --- and $499.59\unit{mrad}$ for matched impedance. \emph{The predictive layer contributes none of this.} Because the matched-gain protocol fixes the first move to the same $(K,D)$, the proposed controller is algebraically an impedance law with a disturbance-observer feedforward, and reproduces it to within $3\times10^{-6}\unit{mrad}$ on every sample of every unconstrained scenario; its constraints are active on $0.0\%$ of solves in the primary benchmark and $0.1\%$ in the near-boundary stress case, and a reactive clip with the same estimator matches it there exactly. Two failure modes appear when the constraints are made to bind. Tightening the torque box to $40$--$20\unit{Nm}$ makes the QP primal-infeasible on $17.7$--$49.7\%$ of solves; its fallback then drives the joint into its mechanical end-stop at up to $8.11\unit{rad/s}$, where the same estimator with a plain clip never reaches the stop at all. And with an explicit two-mass plant and a finite-bandwidth inner torque loop, a $60\unit{Nm}$ \emph{commanded} bound permits $73.0\unit{Nm}$ of \emph{delivered} spring torque, a structural consequence of a fixed closed-loop zero that no simple derating removes. Against a reimplementation of the closest published method \cite{ref37} on that paper's own plant, the estimation and feedforward structure used here reaches $4.84^\circ$/$6.54^\circ$ maximum hip/knee error against $7.27^\circ$/$10.42^\circ$ --- with the caveat that we tuned both, and with the incidental finding that the stated horizon $N_p=5$ is marginally unstable in that reproduction. The estimator is the part of this architecture the evidence supports; the predictive safety layer is not yet earning its place. Scope is stated in Section~\ref{sec:scope}.
\end{abstract}

\begin{IEEEkeywords}
Interaction dynamics, knee rehabilitation, exoskeleton, model predictive control, offset-free MPC, Kalman filter, series elastic actuator, disturbance estimation, impedance control, MuJoCo.
\end{IEEEkeywords}

\section{Introduction}
\IEEEPARstart{P}{hysical} rehabilitation is not only a trajectory-tracking problem. It is an interaction-dynamics problem in which the robot, the actuator, the patient limb, involuntary spasm, voluntary effort, and unmodeled compliance jointly determine the interface motion. The base pHRI framework \cite{ref33} shows that, after appropriate feedforward, such interaction dynamics can be represented by a constant-coefficient double integrator with an estimated persistent disturbance. This paper asks whether that representation survives a rehabilitation-specific actuator and sensing stack: a series-elastic-actuated (SEA) knee exoskeleton.

The knee setting is a useful stress case for the framework. Fixed-gain impedance control is transparent and clinically familiar, but a constant interaction disturbance produces the equilibrium error $e_\infty=\tau_h/K$. A $15\unit{Nm}$ disturbance at $K=30\unit{Nm/rad}$ produces $500\unit{mrad}$ ($28.6^\circ$) error. Increasing stiffness reduces this error but changes the physical interaction and can increase contact load on a recovering joint. A fair evaluation of prediction or disturbance estimation must therefore match the realized stiffness and damping rather than allowing a predictive controller to win by becoming stiffer.

Within the interaction-dynamics view, different robots differ mainly in the recovery map from physical coordinates to the normalized residual model. A rigid manipulator uses computed torque or operational-space feedforward; a hydraulic or cable device uses pressure or tension compensation; a quasi-direct-drive joint uses current-based torque compensation. For the SEA knee, the inner force loop and gravity/inertia/damping feedforward reduce the outer-loop plant to a scalar double integrator, while spring deflection and a dynamic residual provide the interaction-disturbance measurement. The MPC layer is therefore not a knee-specific impedance law but a scalar instantiation of the same normalized interaction MPC used in the base framework.

This paper makes the following evidence-aligned contributions. They are ordered by how well the evidence supports them, and the last two are negative results about the architecture's own predictive layer.
\begin{enumerate}
\item \textbf{An SEA dynamic-residual patient-torque estimator with an explicit steady-state target}, which is what actually removes the interaction offset. It uses noisy finite-differenced acceleration, noisy SEA torque, deliberately mismatched inertia and damping, and no access to the simulated patient torque, and it reaches $0.70\unit{mrad}$ steady-state error and $3.09\unit{mrad}$ full-run RMS, where matched impedance leaves $499.59\unit{mrad}$ and a best-tuned plain integral law leaves $3.26\unit{mrad}$ steady-state but never better than $157\unit{mrad}$ RMS (Section~\ref{sec:whatworks}).
\item \textbf{A closed-loop SEA outer-loop reduction} of the interaction dynamics to the constant double-integrator backbone of \cite{ref33}, with all actuator and limb detail localized to feedforward and disturbance measurement, characterized directly against an explicit two-mass spring plant (Section~\ref{sec:twomass}).
\item \textbf{A matched-gain audit showing that the receding-horizon layer is inert in every scenario reported here.} The matched-impedance protocol fixes the QP's first move to $(K,D)=(30,5)$, so in the unconstrained regime the controller is algebraically impedance plus a disturbance-observer feedforward; it reproduces that law to within $3\times10^{-6}\unit{mrad}$ on every sample, its constraints are active on $0.0$--$0.1\%$ of solves, and a reactive clip with the same estimator matches it on the near-boundary stress case (Section~\ref{sec:whatworks}). We report this rather than presenting a five-controller table in which three entries are the same controller by construction.
\item \textbf{A failure mode of the hard predicted-state constraints.} Tightening the commanded-torque box until it binds makes the QP primal-infeasible on $17.7\%$ ($40\unit{Nm}$) to $49.7\%$ ($20\unit{Nm}$) of solves. The implementation's fallback then becomes the safety controller and drives the joint into its mechanical end-stop at up to $8.11\unit{rad/s}$, while the same estimator behind a plain torque clip never reaches the stop (Section~\ref{sec:hardfail}).
\item \textbf{A commanded-versus-delivered torque gap.} With an explicit two-mass plant and a pole-placed finite-bandwidth inner loop, a $60\unit{Nm}$ commanded bound permits $73.0\unit{Nm}$ of delivered spring torque. The cause is structural --- the inner loop's closed-loop zero sits at $-\omega_0/2$ for any design bandwidth, giving a bandwidth-invariant $\sim43\%$ step overshoot --- and neither natural derating heuristic recovers the bound (Section~\ref{sec:twomass}).
\item \textbf{An external comparison with its caveats stated.} The closest published method, the ESO-based disturbance-rejection MPC of \cite{ref37}, is reimplemented and run head to head against this formulation on that paper's own 2-DOF plant (Section~\ref{sec:external}). Reproducing it also surfaced that its stated horizon $N_p=5$ gives a marginally unstable closed loop here ($\rho=1.0366$, and $>1$ at every control weight and sample period tested), so the reproduction uses $N_p=20$.
\item \textbf{Rehabilitation-specific scheduling and limits}: bounded Assist-as-Needed, a corrective-channel energy tank, and constrained OSQP stress cases are implemented and tested; the controller is additionally run in direct MuJoCo and on a posture-clamped MyoSuite knee slice.
\end{enumerate}

\subsection{Scope}
\label{sec:scope}
Stated once here rather than after each result. The evaluation is single-joint, gravity-compensated simulation under bounded parameter and sensor mismatch; there is no hardware, no patient, and no identification trial. Tracking error and commanded torque are control metrics, not clinical ones: comfort, safety, motor learning, and intent recognition require human subjects and a safety-rated monitor, and the Assist-as-Needed logic demonstrated here is mechanical mode selection under a known simulated torque direction, not validated intent classification. The $15\unit{Nm}$ disturbance magnitude is not anchored to a measured spasm torque --- the Modified Ashworth Scale grades resistance qualitatively and defines no torque-per-grade mapping --- and because a matched-impedance baseline has $e_\infty=\tau_h/K$ exactly, the headline improvement \emph{ratio} is set by that choice; Section~\ref{sec:protocol} places the magnitude against the two scales this study does fix and sweeps it. The energy tank bounds the modeled corrective channel only, not whole-system passivity. Closed-loop stability of the combined estimator-plus-constrained-MPC system is not proved here and is inherited from \cite{ref33}; given the audit above, note that the law actually realized in the unconstrained regime is a static impedance law with a Kalman feedforward, for which that inheritance is a weaker requirement than for the full receding-horizon controller. Constraint satisfaction is established for the \emph{commanded} torque on the tested trajectories only, and Section~\ref{sec:twomass} shows this does not bound the \emph{delivered} torque. No claim of ISO~13482 or IEC~80601-2-78 compliance is made.

\section{Related Work}
\label{sec:relwork}
\subsection{Interaction Dynamics, Impedance, and Predictive Control}
Classical impedance control \cite{ref3} can be interpreted as one static parameterization of desired interaction dynamics. Its transparency and intuitive stiffness/damping parameters explain its prevalence in rehabilitation, but a fixed impedance cannot remove a constant unknown interaction torque without increasing stiffness. Predictive interaction control \cite{ref6} instead regulates the interaction state over a horizon while preserving the impedance interpretation in the unconstrained limit.

Cao, Cheng, and Li \cite{ref7} optimize impedance parameters with an energy tank, while Haninger et al.\ \cite{ref8} co-optimize force reference and impedance changes under learned uncertainty. The present formulation instead optimizes corrective torque for a fixed normalized interaction model. This makes the online problem a quadratic program with constant prediction matrices; the plants, solvers, and hardware differ, so runtimes are not compared directly.

Wang et al.\ \cite{ref9} combine model-predictive impedance control with an extended-state observer. That work and the present controller both estimate lumped disturbances; the distinguishing implementation detail here is the explicit disturbance-compatible steady-state target. Reinforcement learning has also been applied to variable-impedance pHRI \cite{ref18}, but it addresses a different training and certification problem.

The base pHRI framework \cite{ref33} establishes the normalized double-integrator backbone, exact ZOH pair, disturbance augmentation, and impedance interpretation. The present paper specializes them to an SEA knee by deriving the actuator-specific feedforward and measurement channel, then tests whether the same interaction-dynamics observer and target calculation remain effective under rehabilitation-relevant disturbances. Patient torque remains a model-based estimate rather than a direct spring-deflection measurement.

\subsection{Knee Rehabilitation Control}
The Lokomat literature \cite{ref11} illustrates clinically deployed impedance overlays, while Freeman et al.\ \cite{ref12} study iterative learning across repeated rehabilitation trials. Riener et al.\ \cite{ref13} formalize patient-cooperative control. These studies motivate compliant tracking and assistance adaptation, but their hardware, subjects, and protocols are not directly comparable with the present single-joint simulation.

Fang et al.\ \cite{ref27} use sliding-mode control with a perturbation observer under a different disturbance protocol. Because disturbance shape, plant, sensing, and metrics differ, their numerical result is cited for context. Hogan's impedance formulation \cite{ref3}, Pratt and Williamson's SEA concept \cite{ref14}, the personal-care-robot safety standard ISO~13482 \cite{ref10} and the medical-rehabilitation-robot standard IEC~80601-2-78 \cite{ref34}, and the force-control survey \cite{ref15} provide the broader foundations.

Several prior systems already combine model-predictive control, patient-torque estimation, and assist-as-needed logic on a knee or lower-limb exoskeleton, with hardware validation. Caulcrick et al.\ \cite{ref35} estimate human torque from EMG and use MPC to switch among passive, active-assist, and safety modes on a 1-DOF knee exoskeleton, validated on three subjects. Jammeli et al.\ \cite{ref36} combine exact input--output feedback linearization, explicit MPC, a nonlinear disturbance observer for wearer torque, and a Lyapunov stability analysis on a knee rehabilitation orthosis, with real-time experiments on three healthy subjects---the closest prior work to the present formulation, with hardware validation and a standalone closed-loop stability proof. Jin and Guo \cite{ref37} propose an extended-state-observer MPC for a lower-limb exoskeleton with a closed-loop stability analysis, validated in ADAMS/Simulink co-simulation rather than hardware. Jaimes et al.\ \cite{ref38} use a disturbance observer for sensorless torque estimation on a lower-limb rehabilitation robot with hardware trials.

The present work's positive contribution differs from \cite{ref35,ref36,ref37,ref38} on four points, and its negative findings (Sections~\ref{sec:whatworks} and~\ref{sec:hardfail}) apply in part to any of them that shares the same hard-constrained predicted-state formulation: an SEA dynamic-residual torque estimate rather than EMG \cite{ref35} or a generalized-momentum/extended-state observer \cite{ref36,ref37,ref38}; an explicit offset-free steady-state target computed from the estimated disturbance (Section~IV-C), distinct from the feedback-linearized disturbance-observer correction of \cite{ref36} and the extended-state estimate of \cite{ref37}; a matched-impedance evaluation protocol that isolates estimation and target-calculation effects from feedback-gain differences (Section~VI-B), which none of \cite{ref35,ref36,ref37,ref38} reports; and the normalized configuration-invariant interaction-dynamics representation of \cite{ref33}, carried here to a single-DOF SEA joint. The contribution is this combination and its matched-impedance evaluation, in single-joint simulation, together with what that evaluation reveals about the predictive layer itself; hardware validation is reported for \cite{ref35,ref36,ref38} and not here. Section~\ref{sec:external} reimplements \cite{ref37} and compares against it directly, rather than positioning by description alone.

\subsection{Sign Conventions}
The following sign conventions, fixed here, are used without repetition throughout the rest of the paper:
\begin{itemize}
\setlength{\itemsep}{2pt}
\item \textbf{$q,\ \dot q$}: positive value denotes knee \textbf{flexion} (angle / velocity), anatomical.
\item \textbf{$\tau_h$}: positive value denotes patient torque in the \textbf{flexion} direction, anatomical (enters the limb Newton equation (2)).
\item \textbf{$d=-\tau_h/\Ieff$}: positive value denotes the disturbance as it enters the error dynamics (9); $d>0\Leftrightarrow$ extension-direction patient torque.
\item \textbf{$\dd$}: Kalman estimate of $d$; $\operatorname{sgn}(\dd)=-\operatorname{sgn}(\tau_h)$ recovers the anatomical direction of patient torque.
\item \textbf{effort sign $\sigeff=\operatorname{sgn}(\tau_h\dot q_d)=-\operatorname{sgn}(\dd\,\dot q_d)$}: positive value denotes \textbf{assist} (torque aids the desired motion); negative denotes \textbf{resist/spasm} (torque opposes it).
\end{itemize}

\section{Interaction Dynamics Formulation for an SEA Knee}

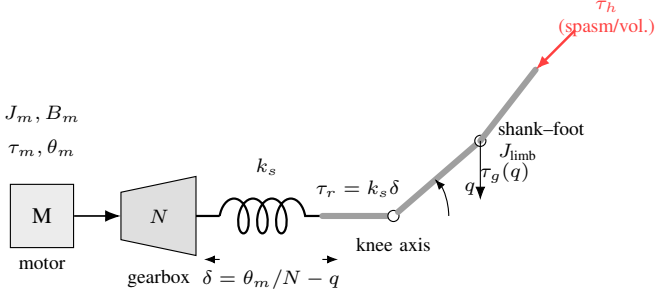
\begin{figure}[t]
\centering
\resizebox{\columnwidth}{!}{%
\begin{tikzpicture}[scale=1.0,every node/.style={font=\footnotesize}]
  \node[draw,fill=gray!15,minimum width=8mm,minimum height=8mm] (mot) at (0,0) {M};
  \node[below=0pt of mot,font=\scriptsize] {motor};
  \node[above,font=\scriptsize] at (0,0.62) {$\tau_m,\theta_m$};
  \node[above,font=\scriptsize] at (0,1.05) {$J_m,B_m$};
  \draw[sig] (mot.east)--(1.0,0);
  \draw[fill=gray!25] (1.0,-0.30)--(1.95,-0.5)--(1.95,0.5)--(1.0,0.30)--cycle;
  \node[font=\scriptsize] at (1.48,0) {$N$};
  \node[below,font=\scriptsize] at (1.48,-0.56) {gearbox};
  \draw[line width=0.8pt] (1.95,0)--(2.25,0);
  \draw[decorate,decoration={coil,aspect=0.55,segment length=2.4mm,amplitude=2.2mm},line width=0.8pt]
        (2.25,0)--(3.55,0);
  \node[above,font=\scriptsize] at (2.85,0.40) {$k_s$};
  \draw[-{Latex[length=1.4mm]}] (2.25,-0.55)--(2.05,-0.55);
  \draw[-{Latex[length=1.4mm]}] (3.55,-0.55)--(3.75,-0.55);
  \node[below,font=\scriptsize] at (2.9,-0.55) {$\delta=\theta_m/N-q$};
  \draw[link] (3.55,0)--(4.45,0);
  \node[above,font=\scriptsize] at (4.0,0.12) {$\tau_r=k_s\delta$};
  \node[joint] (jt) at (4.45,0){};
  \node[below=2pt of jt,font=\scriptsize] {knee axis};
  \draw[link] (jt)--(5.55,0.95);
  \node[joint] at (5.55,0.95){};
  \draw[link] (5.55,0.95)--(6.25,1.85);
  \node[right,font=\scriptsize,align=left] at (5.65,0.95) {shank--foot\\ $J_\text{limb}$};
  \draw[-{Latex[length=1.4mm]}] (5.15,0) arc (0:42:0.7);
  \node[font=\scriptsize] at (5.42,0.34) {$q$};
  \draw[-{Latex[length=1.7mm]}] (5.55,0.95)--(5.55,0.20);
  \node[font=\scriptsize] at (5.86,0.55) {$\tau_g(q)$};
  \draw[-{Latex[length=1.8mm]},red!75,line width=0.9pt] (6.75,2.35)--(6.25,1.85);
  \node[red!75,font=\scriptsize,align=center] at (7.15,2.5) {$\tau_h$\\(spasm/vol.)};
\end{tikzpicture}}
\caption{Series-elastic-actuator (SEA) transmission of the knee joint (the
example platform), mapping the variables of this section onto the physical
parts: motor torque $\tau_m$, rotor angle $\theta_m$, inertia/damping
$J_m,B_m$; gearbox ratio $N$; series-spring stiffness $k_s$ and deflection
$\delta=\theta_m/N-q$; transmitted SEA output torque $\tau_r=k_s\delta$; knee
angle $q$; shank--foot limb inertia $J_\text{limb}$; gravitational moment
$\tau_g(q)$; and the patient interaction torque $\tau_h$ (the disturbance,
involuntary spasm plus voluntary effort). The deflection $\delta$, read from
the motor and joint encoders, provides the SEA output-torque channel
(Section~III-C). Other actuator architectures (spring deflection replaced by motor current, pressure, or cable tension) would replace this recovery stage while leaving the normalized interaction-MPC layer of Section~IV unchanged.}
\label{fig:plant}
\end{figure}

\subsection{SEA-Driven Knee Joint}
\noindent\textit{Motor-side dynamics.}
The exoskeleton actuator (Fig.~\ref{fig:plant}) consists of a brushless motor driving the knee through a gearbox of ratio $N$ with a compliant series spring of stiffness $k_s$ between the gearbox output and the human-interface link. The motor-side equation is
\begin{equation*}
J_m\ddot{\theta}_m + B_m\dot{\theta}_m = \tau_m - \frac{\tau_r}{N} \tag{1}
\end{equation*}
where $\theta_m$ is the motor rotor angle; $J_m$ the combined motor-plus-gearbox inertia; $B_m$ the motor-shaft viscous damping; $\tau_m$ the electromagnetic torque (control input); and $\tau_r = k_s(\theta_m/N - q)$ the spring torque transmitted to the limb.

\noindent\textit{Human-interface dynamics.}
The spring torque drives the combined robot output link and patient shank-foot segment:
\begin{equation*}
\bigl(J_\text{robot} + J_\text{limb}\bigr)\ddot{q} + \bigl(B_\text{robot} + B_\text{limb}\bigr)\dot{q} = \tau_r + \tau_h(t) + \tau_g(q) \tag{2}
\end{equation*}
where $q$ is the knee angle (rad, positive in flexion); $J_\text{limb}\approx 0.446\unit{kg\,m^2}$ is the shank-foot inertia about the knee axis (de~Leva \cite{ref16}, 75~kg / 1.75~m adult: shank $I_\text{shank/knee}=0.181$, foot $I_\text{foot/knee}=0.265\unit{kg\,m^2}$); $\tau_h(t)$ the net patient interaction torque (the disturbance); and $\tau_g(q)=-m_\text{limb}\,g\,L_\text{cm}\cos(\alpha+q)$ the gravitational moment, cancelled by the feedforward of Section~IV-A.

\noindent\textit{Residual single-mass model.}
The exoskeleton closes a high-rate inner force loop on the spring deflection \cite{ref14}, so the outer interaction controller sees the SEA \emph{output torque} directly. With $k_s=200\unit{Nm/rad}$ (joint side), $N=80$, $J_m=1.5\times10^{-4}\unit{kg\,m^2}$, the coupled two-mass resonance is $\approx 26\unit{rad/s}$ ($\approx 4\unit{Hz}$), well above the $\le 0.5\unit{Hz}$ rehabilitation movement band; the inner torque loop (closed-loop bandwidth 20--50~Hz for SEA force control \cite{ref14}) actively damps the spring resonance and presents a near-rigid output torque to the MPC, with any residual two-mass dynamics absorbed by the Kalman disturbance channel of Section~IV-C; this argument is checked directly against an explicit two-mass plant in Section~\ref{sec:twomass} --- which confirms it for nominal tracking and finds where it fails. With the gravity torque $\tau_g(q)$ cancelled by the feedforward, equation~(3) is the \textbf{gravity-compensated residual interaction plant} seen by the MPC layer:
\begin{equation*}
\Ieff\,\ddot{q} + \beff\,\dot{q} = \tau_m^\text{eff} + \tau_h(t) \tag{3}
\end{equation*}
where $\Ieff=J_\text{robot}+J_\text{limb}\approx 0.45\unit{kg\,m^2}$, $\beff=B_\text{robot}+B_\text{limb}\approx 0.50\unit{Nm\,s/rad}$, and $\tau_m^\text{eff}$ is the torque command referred to the joint output. Hardware deployment would require patient-specific identification; no identification trial is performed in this study.

\noindent\textit{Benchmark model and ROM.}
The validation uses the gravity-compensated scalar plant (3) and a custom MuJoCo \cite{ref22} knee-pendulum model (\texttt{knee\_rehab\_1dof.xml}): a fixed thigh, shank-foot body, hinge knee, and direct-torque actuator representing the closed SEA inner loop. MuJoCo computes an initial inertia of $0.4328\unit{kg\,m^2}$, providing a plant distinct from both the analytical simulation and controller model.
The normal knee flexion--extension range is $q\in[0^\circ,120^\circ]=[0,2.094]\unit{rad}$. Redundant hardware end-stops plus an independent software ROM limit are standard practice for powered rehabilitation exoskeletons and are required by the applicable medical-robot standard, IEC~80601-2-78 \cite{ref34}; ISO~13482 \cite{ref10} covers personal-care robots and explicitly excludes rehabilitation medical devices. The MPC ROM constraints (Section~V-A) provide the software limit with a predictive margin, not certified compliance with either standard.

\subsection{Patient Interaction Torque Model}
The patient interaction torque has two physically distinct components:
\begin{equation*}
\tau_h(t) = \underbrace{\tau_\text{spasm}(t)}_{\text{involuntary}} + \underbrace{\tau_\text{vol}(t)}_{\text{voluntary}} \tag{4}
\end{equation*}

\emph{Spasm torque} $\tau_\text{spasm}(t)$: an impulsive or step-like contraction triggered by the stretch reflex, hyperactive (2--5$\times$ normal gain) in neurological patients, typically opposing the robot's trajectory. Simulation model: step onset $\tau_\text{spasm}=15\unit{Nm}$ lasting 1.5~s, a repeatable stress magnitude motivated by, but not a validated equivalent of, a clinically resistive spasm; the Modified Ashworth Scale grades resistance qualitatively and does not define a fixed torque-to-grade mapping, so no specific MAS grade is claimed.

\emph{Voluntary torque} $\tau_\text{vol}(t)$: the net muscular output as the patient follows the prescribed trajectory; it can aid or oppose the motion. The lumped disturbance $d(t)=-\tau_h(t)/\Ieff$ can therefore have either sign---the key rehabilitation-specific complexity: combining $\dd$ with the known desired-motion direction distinguishes AAN assist-reduction from spasm rejection (Section~V-C). All sign conventions follow Section~II-C; the anatomical sign of $\dd$ and the motion-relative effort sign $\sigeff$ coincide only for a fixed motion direction, so the AAN logic uses $\sigeff$, which requires only $\dd$ and the reference velocity $\dot q_d$.

\subsection{Patient-Torque Estimation from the SEA Dynamic Residual}
The spring deflection is a direct measurement of SEA output torque, not patient torque. The deflection $\delta=\theta_m/N-q$ is measured from motor and joint encoders, giving
\begin{equation*}
\tau_r = k_s\,\delta = k_s\!\left(\frac{\theta_m}{N}-q\right) \tag{5}
\end{equation*}
Rearranging the residual joint dynamics gives the patient-torque estimate
\begin{equation*}
\hat{\tau}_h=\hat I_\mathrm{eff}\hat{\ddot q}+
\hat b_\mathrm{eff}\dot q-\tau_r-\hat\tau_g(q). \tag{6}
\end{equation*}
Acceleration is estimated by finite-differencing joint velocity. The validation deliberately uses $\hat I_\mathrm{eff}=0.4725\unit{kg\,m^2}$ and $\hat b_\mathrm{eff}=0.55\unit{Nm\,s/rad}$ while the analytical plant uses $0.45\unit{kg\,m^2}$ and $0.50\unit{Nm\,s/rad}$. It adds $0.35\unit{rad/s^2}$ acceleration noise, $0.50\unit{Nm}$ SEA-torque noise, and randomized torque bias. Thus the estimator is not given $\tau_h$ and the evaluation avoids an inverse crime. The quasi-static balance is a limiting interpretation of (6), not the implemented estimator.

In joint space, the Kalman disturbance state is $\dd=-\hat\tau_h/\hat I_\mathrm{eff}$, distinct from the true physical $d=-\tau_h/\Ieff$ of Section~II-C. The Kalman measurement entry $-\hat I_\mathrm{eff}$ in~(12) maps this acceleration disturbance back to the dynamic-residual torque measurement. The same outer-loop structure can be used with other actuator technologies if an output-torque estimate is available, but that portability is not tested here.

\section{Interaction Dynamics MPC Design}
\subsection{Architecture Overview}
The controller has two layers.

\textbf{Layer 1 --- Feedforward.} An algebraic term cancels the known SEA plant dynamics (effective inertia, damping, gravity), leaving a scalar double integrator as the residual plant:
\begin{equation*}
\tau_\text{ff} = \hat I_\mathrm{eff}\,\ddot{q}_d + \hat b_\mathrm{eff}\,\dot{q} - \tau_g(q) \tag{7}
\end{equation*}
The $-\tau_g(q)$ sign is fixed by~(2): $\tau_g(q)$ enters the plant with a $+$ sign, so cancelling it requires the feedforward to supply $-\tau_g(q)$, consistent with the residual-torque identity~(6). The validation plant is gravity-free ($\tau_g\equiv0$ throughout, per the single-axis, gravity-compensated scope of Section~I), so this sign has no effect on any reported numeric result; it matters only if~(7) is instantiated on a plant with nonzero $\tau_g(q)$. The controller builds the feedforward from its own identified parameters $\hat I_\mathrm{eff}$, $\hat b_\mathrm{eff}$ (Table~\ref{tab:parameters}), not the true plant values $\Ieff$, $\beff$ of~(3); the resulting mismatch is absorbed into $d(t)$ below, alongside patient torque. The inertia term uses the reference acceleration $\ddot{q}_d$; the damping term uses the \emph{actual} velocity $\dot{q}$, so $\hat b_\mathrm{eff}\dot{q}$ cancels the nominal damping and leaves $A_c$ a constant double integrator. In hardware $\dot{q}$ is low-pass filtered against encoder noise; the small filter lag is not added to the prediction model but lumped into $d(t)$ and absorbed by the Kalman state \cite{ref33}. Only three offline-identified parameters ($\hat I_\mathrm{eff}$, $\hat b_\mathrm{eff}$, $m_\text{limb}L_\text{cm}$) are required, keeping the approach demonstration-free.

\textbf{Layer 2 --- Interaction Dynamics MPC.} A receding-horizon QP computes a corrective torque $\tau_\text{mpc}$ on the residual double integrator to minimize tracking error; ROM, torque, and velocity limits can be included as linear inequalities over the $N$-step prediction. The same normalized prediction object would be used for a hydraulic, cable, or QDD knee after replacing the feedforward and measurement channel.

\subsection{Scalar Error Dynamics}
Define $e=q_d-q$ and decompose $\tau=\tau_\text{ff}+\tau_\text{mpc}$. After applying (7) and substituting into (3), the residual error dynamics are
\begin{equation*}
\ddot{e} = -\Gamma_e\,\tau_\text{mpc} + d(t) \tag{8}
\end{equation*}
where $d(t)$ aggregates patient torque, the feedforward mismatch between $(\hat I_\mathrm{eff},\hat b_\mathrm{eff})$ and the true $(\Ieff,\beff)$, and any other unmodeled term, normalized to acceleration units; on the true plant the patient-torque component alone enters as $-\tau_h/\Ieff$. $\Gamma_e=1/\hat I_\mathrm{eff}=2.116\,\mathrm{rad\,s^{-2}\,Nm^{-1}}$ is the controller's own nominal gain, built from $\hat I_\mathrm{eff}$ rather than the true $\Ieff$, since $B_c$ is the controller's model of its own effect on the plant, not a property of the physical plant. Substituting the true plant dynamics (3) and the feedforward (7) makes the input-gain term explicit:
\begin{equation*}
d(t) = -\frac{\tau_h(t)}{\Ieff} + d_\text{ff}(t) + \left(\frac{1}{\hat I_\mathrm{eff}}-\frac{1}{\Ieff}\right)\tau_\text{mpc}(t),
\end{equation*}
where $d_\text{ff}$ collects the feedforward-mismatch terms in $\ddot q_d$ and $\dot q$ and the last, \emph{input-dependent} term is exactly the gap between the controller's gain $\Gamma_e$ and the plant's true input gain $1/\Ieff$. Because this term depends on $\tau_\text{mpc}$ itself, $d(t)$ is not strictly an exogenous disturbance; the Kalman filter's random-walk model treats it as one, a practical approximation valid while $\tau_\text{mpc}$ and the mismatch stay small and slowly varying relative to the estimator bandwidth, not an exact augmented-system derivation. The state $x_e=[e,\dot{e}]^\top$ satisfies
\begin{equation*}
\dot{x}_e = \underbrace{\begin{bmatrix}0 & 1 \\ 0 & 0\end{bmatrix}}_{A_c} x_e + \underbrace{\begin{bmatrix}0 \\ -\Gamma_e\end{bmatrix}}_{B_c}\tau_\text{mpc} + \underbrace{\begin{bmatrix}0 \\ 1\end{bmatrix}}_{G_c} d(t) \tag{9}
\end{equation*}
The disturbance enters through $G_c=[0,1]^\top$, distinct from the control matrix $B_c$. As in \cite{ref33}, $A_c$ is nilpotent, so the discrete $A_d$ is exact and configuration-independent with closed-form ZOH matrices:
\begin{equation*}
A_d=\begin{bmatrix}1 & \Delta t \\ 0 & 1\end{bmatrix},\ B_d=\begin{bmatrix}-\tfrac{\Delta t^2}{2}\Gamma_e \\ -\Delta t\,\Gamma_e\end{bmatrix},\ G_d=\begin{bmatrix}+\tfrac{\Delta t^2}{2} \\ +\Delta t\end{bmatrix} \tag{10}
\end{equation*}
($G_d=-B_d/\Gamma_e$: positive, since the disturbance enters the error directly.) The constant matrices permit offline construction of the prediction and Hessian matrices. The validation uses $N=20$ at both update rates.

\subsection{Disturbance Augmentation for Offset-Free Tracking}
Following offset-free MPC theory \cite{ref20}, the error state is augmented with a random-walk disturbance state $\dd\in\mathbb{R}$ estimated by a steady-state Kalman filter \cite{ref19}:
\begin{equation*}
\begin{bmatrix}x_e(k+1) \\ \dd(k+1)\end{bmatrix} = \underbrace{\begin{bmatrix}A_d & G_d \\ 0 & 1\end{bmatrix}}_{\mathcal{A}}\begin{bmatrix}x_e(k) \\ \dd(k)\end{bmatrix} + \begin{bmatrix}B_d \\ 0\end{bmatrix}\tau_\text{mpc}(k) \tag{11}
\end{equation*}
A steady-state Kalman filter estimates $\dd(k)$ from three measurements:
\begin{equation*}
y_k = \begin{bmatrix}1 & 0 & 0 \\ 0 & 1 & 0 \\ 0 & 0 & -\hat I_\mathrm{eff}\end{bmatrix}\begin{bmatrix}e \\ \dot{e} \\ \dd\end{bmatrix}_k + v_k \tag{12}
\end{equation*}
The third measurement is the dynamic residual (6), $\hat\tau_h\approx-\hat I_\mathrm{eff}\dd$. Offset-free operation additionally requires a disturbance-compatible target. For constant $\dd$, zero error acceleration requires
\begin{equation*}
\tau_\mathrm{ss}=\hat I_\mathrm{eff}\dd=-\hat\tau_h. \tag{13}
\label{eq:target}
\end{equation*}
The optimizer therefore uses $\tau_\mathrm{mpc}=\tau_\mathrm{ss}+v$ and regulates the deviation sequence $V$. This target calculation, rather than disturbance propagation alone, removes the constant offset. The nominal guarantee assumes observability, estimator convergence, an accurate constant-disturbance model, and QP feasibility; the randomized mismatch results quantify non-ideal residual error.

\emph{Remark (Bidirectional disturbance).} Patient torque can have either sign. Its product with desired velocity indicates whether torque aids or opposes the commanded motion, but sign alone does not establish clinical intent. The present paper reports disturbance rejection and torque crediting only; validated Assist-as-Needed classification requires a separate human-subject protocol.

\subsection{Receding-Horizon QP}
\textbf{Prediction stack.} Applying (10) recursively and stacking the $N$ predicted error states gives
\begin{equation*}
\Phi=\begin{bmatrix}A_d \\ A_d^2 \\ \vdots \\ A_d^N\end{bmatrix}\in\mathbb{R}^{2N\times 2},\quad \Gamma=\begin{bmatrix}B_d & & \\ A_d B_d & B_d & \\ \vdots & \ddots & \\ A_d^{N-1}B_d & \cdots & B_d\end{bmatrix}.
\end{equation*}
Because $A_d$ is constant, $\Phi$ and $\Gamma$ are precomputed offline. Let $V=[v_0,\ldots,v_{N-1}]^\top$ denote deviations from $\tau_\mathrm{ss}$. In the nominal model, the constant target and estimated disturbance cancel, giving $Y=\Phi x_0+\Gamma V$. The condensed cost is
\begin{equation*}
H=\Gamma^\top\bar{Q}\Gamma+\bar{R},\qquad f=\Gamma^\top\bar{Q}\Phi x_0 \tag{14}
\end{equation*}
with $\bar{Q}=\operatorname{blkdiag}(Q,\ldots,Q,Q_f)$, $\bar{R}=R_u I_N$; both $H$ and $H^{-1}$ are precomputed offline. At each step the condensed QP is
\begin{equation*}
\min_{V}\;\tfrac{1}{2}V^\top H\,V + f^\top V \tag{15}
\end{equation*}
subject to
\begin{equation*}
\begin{aligned}
\tau_\text{ff}(k)+\tau_\text{ss}+v(k) &\in [-\taumax,\taumax] && \text{(cmd.-torque limit)} \\
q_d(k)-e(k) &\in [q_\text{min},q_\text{max}] && \text{(ROM constraint)} \\
|\dot{q}_d(k)-\dot{e}(k)| &\le \omega_\text{max} && \text{(velocity limit)}
\end{aligned} \tag{16}
\end{equation*}
The current-tick feedforward $\tau_\text{ff}(0)$ is~(7), evaluated at the
measured $\dot q$; for $k\ge1$, $\dot q$ is not yet measured, so the
horizon row instead predicts it from the same $\dot e(k)$ used by the
velocity row above:
\begin{equation*}
\tau_\text{ff}(k)=\hat I_\mathrm{eff}\,\ddot{q}_d(k)+\hat b_\mathrm{eff}\bigl(\dot{q}_d(k)-\dot{e}(k)\bigr),\qquad k\ge1.
\end{equation*}
Because $\dot e(k)$ is itself a linear function of $V$ through the
prediction stack, this makes the torque row affine in $V$, on the same
footing as the ROM and velocity rows; the implementation folds the
$\hat b_\mathrm{eff}\dot e(k)$ term into the row's coefficient on $V$
rather than into its bound, with $\taumax=60\unit{Nm}$, $q_\text{min}=0$, $q_\text{max}=2.094\unit{rad}$, and $\omega_\text{max}=2.0\unit{rad/s}$. The primary matched-gain study uses the unconstrained closed form; a separate OSQP stress test executes all three inequality sets. This establishes software enforcement of the \emph{commanded}-torque bound on the tested trajectory, not recursive feasibility, safety certification, or a bound on the \emph{delivered} SEA output torque under finite inner-loop bandwidth (Section~VIII).

\subsection{Equivalence to Classical Impedance}
In the unconstrained, disturbance-free limit the infinite-horizon MPC is a static LQR feedback rendering the classical joint-impedance law \cite{ref33}:
\begin{equation*}
\tau^\text{cmd} = \hat I_\mathrm{eff}\,\ddot{q}_d + \hat b_\mathrm{eff}\,\dot{q} - \tau_g(q) + K_d\,e + D_d\,\dot{e} \tag{17}
\end{equation*}
with realized stiffness/damping $(K_d,D_d)$ given by the first-input Riccati/finite-horizon gain, not by the entries of $Q$ themselves. The validation tunes separate weight triples at 100 and 500~Hz so both realize exactly $(30,5)$. This matching is essential: otherwise rate-dependent gain changes can be mistaken for predictive disturbance rejection.

\section{Rehabilitation-Specific Extensions}
\subsection{ROM Constraint Linearization in QP}
The state constraint $q_\text{min}\le q(k)\le q_\text{max}$ is expressed in the error state via $q(k)=q_d(k)-e(k)$:
\begin{align}
q_d(k)-e(k)\ge q_\text{min} &\Rightarrow e(k)\le q_d(k)-q_\text{min} \tag{18a}\\
q_d(k)-e(k)\le q_\text{max} &\Rightarrow e(k)\ge q_d(k)-q_\text{max} \tag{18b}
\end{align}
Since $e(k)$ is linear in the decision sequence, the ROM bounds add two rows per prediction step. The OSQP implementation also includes total-torque and predicted-velocity rows. On the reported stress trajectory it remains feasible and satisfies the tested bounds. A terminal invariant set, backup controller, and proof of recursive feasibility are not included.

\subsection{Passivity-Oriented Corrective-Channel Energy Tank}
The implemented energy tank limits energy injected by the corrective outer-loop torque:
\begin{equation*}
\begin{aligned}
p_k&=\tau_{\mathrm{mpc},k}\dot q_k,\\
T_{k+1}&=\operatorname{clip}_{[0,T_{\max}]}\!
\left(T_k-[p_k]_+\Delta t+\eta[-p_k]_+\Delta t\right).
\end{aligned}\tag{19}
\end{equation*}
At each integration step, positive corrective power $\tau_{\mathrm{mpc}}\dot q$ withdraws energy from $T_k$; power absorbed by the corrective channel replenishes the tank with efficiency $\eta=0.8$, up to a finite capacity. If a requested command would make $T_{k+1}<0$, the torque is scaled to use exactly the available energy. The implementation therefore guarantees $T_k\ge0$ for the modeled corrective channel by construction.

This is a \emph{passivity-oriented channel result}, not a passivity certificate for the complete robot--human system. Model-based feedforward, the inner SEA torque loop, communication delay, and physical damping are outside the tank accounting. The depleted-tank stress test intentionally demonstrates the safety/performance tradeoff: energy remains nonnegative and 2441 interventions occur, while RMS tracking error rises to $368.7\unit{mrad}$.

\subsection{Motion-Relative Torque Sign and Assist-as-Needed Scope}
The motion-relative effort sign $\sigeff$ (Section~II-C) encodes whether the patient's torque aids or opposes the desired motion:
\begin{equation*}
\begin{aligned}
\sigeff(t)
&=\operatorname{sgn}(\tau_h\dot q_d)
 =-\operatorname{sgn}(\dd\,\dot q_d)\\
&=\begin{cases}
+ & \text{torque aids motion},\\
- & \text{torque opposes (spasm)} .
\end{cases}
\end{aligned} \tag{20}
\end{equation*}
This quantity requires only $\dd$ and the known reference velocity. It identifies whether estimated patient torque is mechanically aligned with commanded motion, but cannot by itself distinguish voluntary intent, reflex, co-contraction, or alignment artifacts. The implemented compliant schedule is
\begin{equation*}
K_d(t)=\max\bigl(K_\text{min},\,K_\text{nominal}-\alpha_\text{AAN}\,|\dd|\bigr) \tag{21}
\end{equation*}
with verified modes $(K,D)=(30,5)$ and $(10,5)$. When assistance is detected, cancellation credit is bounded by
\begin{equation*}
|\tau_\mathrm{credit}|\le K_\mathrm{soft}e_\mathrm{budget},
\qquad e_\mathrm{budget}=60\unit{mrad}. \tag{22}
\end{equation*}
Bounding the credit keeps the tracking deviation within the stated budget even when assistance is detected. The demonstrated claim is mechanical Assist-as-Needed scheduling under known simulated torque direction, not clinical intent recognition.

\section{Evidence-Aligned Validation}
\subsection{Protocol and Fair-Comparison Design}
\label{sec:protocol}
The reference trajectory is
\begin{equation*}
q_d(t)=\frac{\pi}{3}+r(t)\frac{\pi}{6}\sin\left(\frac{\pi}{2}t\right),
\qquad r(t)=\min(1,t). \tag{23}
\label{eq:reference}
\end{equation*}
A $15\unit{Nm}$ patient torque opposes the instantaneous desired-motion direction from 1.5 to 3.0~s in each 4~s cycle, and four cycles are simulated. Defining the disturbance relative to desired motion avoids treating the same anatomical torque as assistance on one half-cycle and resistance on the other.

The analytical plant uses $I_\mathrm{eff}=0.45\unit{kg\,m^2}$ and $b_\mathrm{eff}=0.50\unit{Nm\,s/rad}$, while the controller and dynamic-residual estimator use $0.4725\unit{kg\,m^2}$ and $0.55\unit{Nm\,s/rad}$. Acceleration and SEA-torque noise are injected before the Kalman update. The prediction horizon is $N=20$ at both 100 and 500~Hz, corresponding to physical horizons of 200 and 40~ms.

Five controllers are compared:
\begin{enumerate}
\item classical impedance at $(K,D)=(30,5)$;
\item MPC without disturbance estimation at 100~Hz;
\item offset-free MPC with Kalman estimation at 100~Hz;
\item MPC without disturbance estimation at 500~Hz; and
\item offset-free MPC with Kalman estimation at 500~Hz.
\end{enumerate}
The cost weights are tuned offline so the first unconstrained MPC move realizes exactly $K=30\unit{Nm/rad}$ and $D=5\unit{Nm\,s/rad}$ at both rates. Automated tests verify these gains to numerical tolerance, so gain differences between controllers cannot confound the comparison.

Metrics are full-run RMS error, RMS and peak error during disturbance windows, and mean absolute error over the final 0.3~s of each disturbance window. These are simulation metrics, not clinical accuracy claims.

\noindent\textit{What sets the headline ratio.} A matched-impedance baseline satisfies $e_\infty=\tau_h/K$ exactly, so the size of the improvement reported below is fixed by the chosen disturbance magnitude rather than measured. Sweeping $\tau_h$ over $3$--$40\unit{Nm}$ confirms this to within integration error: the classical controller's steady-state error is $100.04$, $199.93$, $333.11$, $499.59$, $832.55$ and $1335.14\unit{mrad}$ against a predicted $\tau_h/K$ of $100.0$, $200.0$, $333.3$, $500.0$, $833.3$ and $1333.3\unit{mrad}$, while the offset-free controller stays between $0.61$ and $0.89\unit{mrad}$ throughout. The offset-free result is therefore essentially independent of $\tau_h$, and quoting a ratio ($\sim700\times$ at $15\unit{Nm}$, $\sim160\times$ at $3\unit{Nm}$) says more about the benchmark than the controller; the absolute residual is the meaningful number.

The magnitude itself cannot be anchored to a measured spasm torque (Section~\ref{sec:scope}), but it can be placed against two scales this study already fixes. Using the same de~Leva anthropometry \cite{ref16} that sets $J_\text{limb}$, the patient's shank--foot segment weighs $4.28\unit{kg}$ with its combined centre of mass $0.261\unit{m}$ from the knee axis, so its peak gravitational moment is $11.0\unit{Nm}$: the benchmark disturbance is $1.37\times$ that, and $25\%$ of the device's own $60\unit{Nm}$ torque limit. It is a stress magnitude of the same order as holding the limb against gravity, which is a defensible design case without being a clinical one.

\subsection{Matched-Impedance Results}
\begin{table}[t]
\caption{Matched-Impedance Results and Layer Attribution (Measured, $K=30$, $D=5$)}
\label{tab:matched}
\centering
\footnotesize
\setlength{\tabcolsep}{4pt}
\begin{tabular}{lrrrr}
\toprule
Controller & RMS & Dist. RMS & Peak & SS\\
 & \multicolumn{4}{c}{(mrad)}\\
\midrule
C1 classical impedance & 293.77 & 464.16 & 527.12 & 499.59\\
C2 MPC 100, no est. & 294.04 & 465.08 & 527.47 & 499.69\\
C4 MPC 500, no est. & 293.77 & 464.16 & 527.12 & 499.59\\
\midrule
C3 MPC $+$ Kalman 100 & 5.63 & 7.99 & 22.85 & 1.17\\
C5 MPC $+$ Kalman 500 & \textbf{3.09} & \textbf{2.53} & \textbf{7.27} & \textbf{0.70}\\
\midrule
\multicolumn{5}{l}{\emph{Added here to attribute the improvement:}}\\
B0 impedance $+$ integral, best SS & 183.49 & 211.55 & 427.13 & 3.26\\
\ \ \ \ \emph{($K_i=70$)} & & & & \\
B0$'$ same, best RMS ($K_i=140$) & 157.15 & 176.45 & 383.66 & 51.53\\
B1 impedance $+$ est.\ $+$ target & 3.09 & 2.53 & 7.27 & 0.70\\
\ \ \ \ \emph{(no horizon, no QP)} & & & & \\
\bottomrule
\end{tabular}
\end{table}

Table~\ref{tab:matched} confirms the equilibrium physics, and it should be read as an identity rather than a comparison. Classical impedance and both no-estimator MPC controllers produce approximately $500\unit{mrad}$ steady-state error, matching $\tau_h/K$; they agree because the matched-gain protocol makes them the same controller, which is a check on the implementation, not a result. Three of the five rows therefore carry one piece of information between them.

The combination of dynamic-residual estimation and the target input \eqref{eq:target} changes the result. At 100~Hz, steady-state error falls to $1.17\unit{mrad}$ and peak error to $22.85\unit{mrad}$. At 500~Hz, the corresponding values are $0.70$ and $7.27\unit{mrad}$. The rate mainly improves estimator and sample-and-hold transients; the offset-free mechanism is already effective at 100~Hz. What Table~\ref{tab:matched} does \emph{not} establish is which layer of the architecture produces that change, and Section~\ref{sec:whatworks} shows the answer is not the one the controller is named after.

\begin{figure*}[t]
\centering
\includegraphics[width=0.94\textwidth]{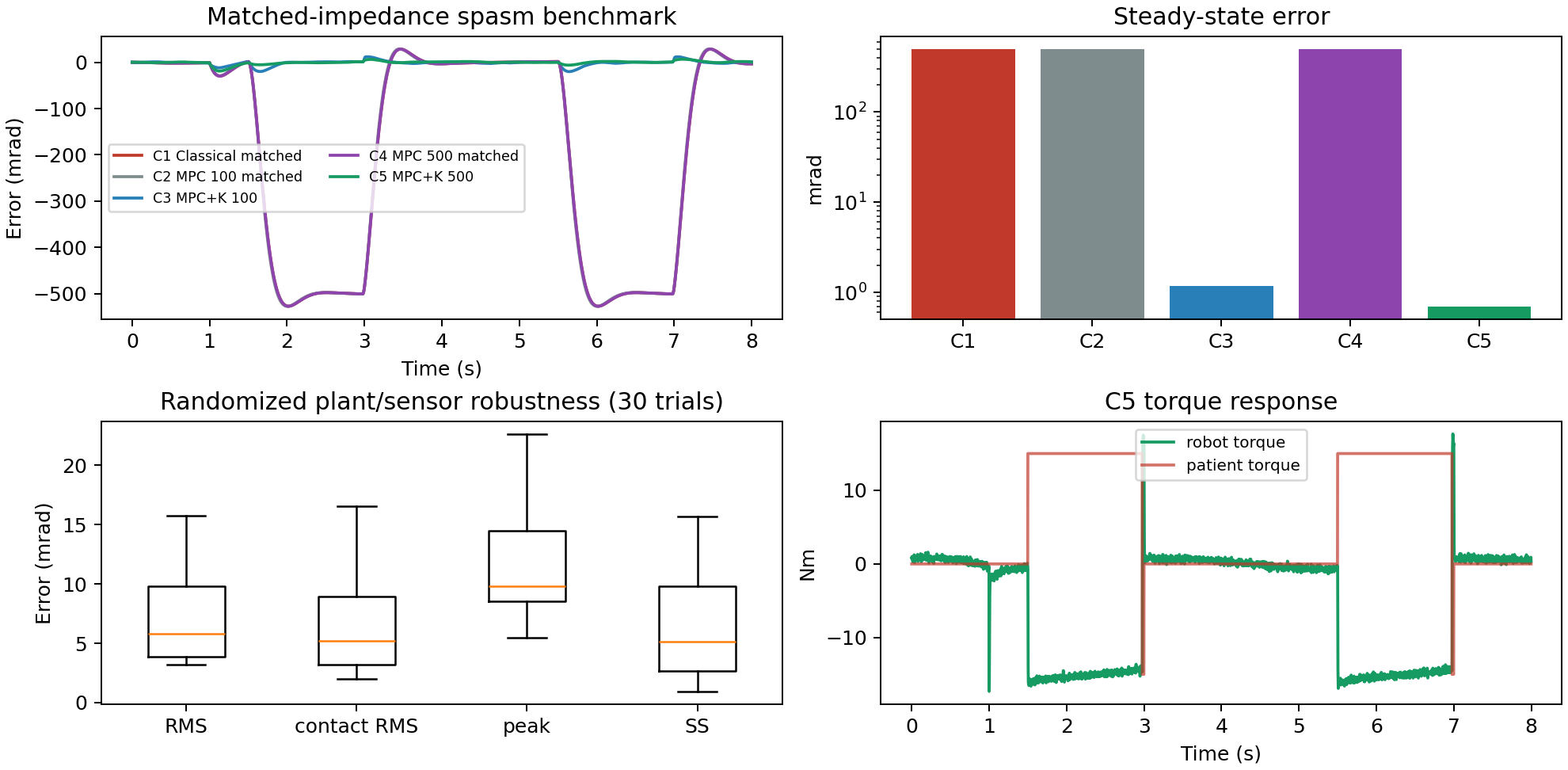}
\caption{Evidence-aligned validation. Top left: tracking errors for stiffness- and damping-matched controllers. Top right: steady-state error. Bottom left: 30 randomized plant and sensor trials, with the dotted 87~mrad line a representative engineering design tolerance chosen for this benchmark, not a validated patient-specific clinical threshold. Bottom right: patient and robot torque for the 500~Hz offset-free controller.}
\label{fig:validation}
\end{figure*}

\subsection{Which Layer Produces the Result?}
\label{sec:whatworks}
Table~\ref{tab:matched} contains three cells of a natural $2\times2$: static feedback versus receding horizon, crossed with no estimator versus estimator-plus-target. The fourth cell --- classical impedance carrying the \emph{same} Kalman filter, the same dynamic residual, and the same steady-state target $\tau_\mathrm{ss}=\hat I_\mathrm{eff}\dd$, but with no horizon, no prediction matrices and no QP --- is the one that identifies the mechanism, and it is the cell the original protocol omits. The last two rows of Table~\ref{tab:matched} fill it in, with a plain integral-action law alongside for reference.

Three readings follow.

\noindent\textit{The estimator and target are doing the work, but the margin is transient, not steady-state.} Adding a plain integral term to the matched impedance law --- the textbook way to remove a constant offset, with no estimator and no SEA residual --- gives a much stronger baseline than matched impedance alone, and it has to be tuned properly rather than dismissed. Swept over $K_i\in[5,300]\unit{Nm\,rad^{-1}s^{-1}}$, steady-state error is minimised at $K_i=70$ ($3.26\unit{mrad}$) and full-run RMS at $K_i=140$ ($157.15\unit{mrad}$), and no gain does well on both. Against the better of those, the estimator's \emph{steady-state} advantage is a factor of $4.7$ ($0.70$ vs.\ $3.26\unit{mrad}$): real, but not the order of magnitude a coarser sweep would suggest.

The durable advantage is in the transient. Integral action must accumulate the error at every spasm onset before it acts, so its full-run RMS never falls below $157\unit{mrad}$ at any gain, whereas the dynamic residual observes the disturbance within one estimator time constant ($6\unit{ms}$ rise to 90\% at 500~Hz) and holds $3.09\unit{mrad}$ --- a factor of $51$ against the best-RMS integral gain. On a protocol with four disturbance onsets in $16\unit{s}$, that transient term is most of the error a patient would experience. This is the contribution, and it belongs to the measurement channel, not to the optimizer.

\noindent\textit{The horizon is inert.} Rows C5 and B1 of Table~\ref{tab:matched} are not merely close; they agree to every reported digit, and the largest per-sample difference in tracking error over the whole $16\unit{s}$ run is $2.6\times10^{-6}\unit{mrad}$ --- floating-point path difference between solving the QP and evaluating the gain --- against a $528.7\unit{mrad}$ maximum difference between C1 and C5. The same holds in the assistive mode ($2.54/1.30/3.77/0.59\unit{mrad}$ for both). This is not a numerical coincidence but a consequence of the fair-comparison protocol itself: the matched-gain construction fixes $-(H^{-1}\Gamma^\top\bar Q\Phi)_0=[K,D]$, and in the nominal model the disturbance and target columns cancel by the offset-free property, so
\begin{equation*}
\tau_\mathrm{mpc}=\tau_\mathrm{ss}+v_0=\hat I_\mathrm{eff}\dd+K e+D\dot e \tag{24}
\end{equation*}
exactly --- an impedance law with a disturbance-observer feedforward. The protocol that makes the comparison fair also makes the horizon unobservable in the result.

\noindent\textit{The constraints almost never bind.} A horizon can still matter through active constraints, so we counted them. Over the standard protocol with the constrained solver enabled, the commanded-torque, ROM and velocity rows are active on $0.0\%$, $0.0\%$ and $0.0\%$ of QP solves respectively (nonzero OSQP dual, tolerance $10^{-9}$). On the deliberately hard near-boundary case of Section~\ref{sec:nearboundary} the torque row is active on $0.1\%$ of solves and the ROM and velocity rows on none. Moreover, repeating that case with the reactive middle term --- matched impedance, the same estimator, a plain torque clip, no predicted-state rows --- gives an identical outcome: peak commanded torque $60.00\unit{Nm}$ for both, maximum joint angle $1.966\unit{rad}$ for both, and neither reaches the ROM ceiling, while the estimator-free reactive baseline used previously peaks at $42.32\unit{Nm}$ and arrives at the stop at $5.72\unit{rad/s}$. The advantage previously attributed to prediction in that scenario is the estimator's: it cancels the $40\unit{Nm}$ push, which is what frees the torque budget.

We report this because a five-controller table in which three rows are the same controller and the remaining difference is reproduced exactly by a static law is not evidence for model-predictive control. It is evidence for the estimator. Section~\ref{sec:external} adds the external comparison these internal ablations cannot provide. (\texttt{simulation/validate\_mpc\_contribution.py}.)

\subsection{An External Comparison, With Its Caveats Stated}
\label{sec:external}
The comparisons above are all internal, which is a real limitation: they establish which of \emph{our} layers does the work, not how the whole controller stands against a published method. We therefore reimplemented the closest one --- the extended-state-observer disturbance-rejection MPC of Jin and Guo \cite{ref37} --- and ran it head to head against this paper's error-state formulation on a shared plant. It is reported here with three caveats attached, because a reimplementation that we tuned, on a plant we wrote, is weak evidence about someone else's method and strong evidence only about the comparison being possible at all.

\emph{Plant.} The comparison runs on \cite{ref37}'s own 2-DOF sagittal double-pendulum exoskeleton (hip and knee, Table~1 parameters of that paper) at $T_s=10\unit{ms}$, not on the SEA knee of the rest of this paper. It therefore says nothing about the SEA reduction of Section~\ref{sec:twomass}; it compares controller structures on their plant.

\emph{Result.} After a $1\unit{s}$ settling window, the reproduction of \cite{ref37} (original-state velocity-form MPC with the linear ESO estimating unmeasured velocities and the lumped disturbance) reaches maximum tracking errors of $7.27^\circ$ at the hip and $10.42^\circ$ at the knee (mean $2.58^\circ$/$3.32^\circ$), against $4.84^\circ$ and $6.54^\circ$ (mean $1.85^\circ$/$1.92^\circ$) for a computed-torque feedforward plus per-joint error-state double-integrator MPC with the integrating-disturbance Kalman filter used here.

\emph{First caveat: our tuning.} Both controllers were tuned by us on our implementation of their plant. A reimplementation measures the reimplementer's tuning at least as much as the method, which is why we did not present this as a ranking in the abstract or the contributions.

\emph{Second caveat, and a finding in its own right: the stated horizon is not stable here.} Reference~\cite{ref37} specifies $N_p=5$, $N_c=2$. In this reproduction the linearized velocity-form closed loop --- $\xi^+=(\mathcal A_a-\mathcal B_a M_1F)\xi$, with perfect state feedback and the exogenous terms dropped --- has spectral radius $\rho=1.0366$ at the control weight used, falling only to $1.0041$ at a weight $10^3$ times larger: greater than one at \emph{every} weight tested, and greater than one across $T_s\in[2,20]\unit{ms}$ and across the hip--knee coupling angle. It is marginally unstable without a terminal cost. We therefore used $N_p=20$, for which $\rho=0.9316$. This is a property of \emph{this} reproduction --- forward-Euler discretization, no terminal cost, perfect state feedback --- and not a proof about the original controller, which may differ in any of those; at $T_s=50\unit{ms}$ the same audit returns $\rho=0.8117$, i.e. stable. It is reported because a reader reproducing \cite{ref37} at the stated horizon should expect to meet it. (\texttt{simulation/jin\_guo\_eso\_mpc.py}, \texttt{horizon\_audit()}.)

\emph{Third caveat: what it does not settle.} Neither controller here is the constrained one. The comparison is between an ESO-plus-velocity-form-MPC structure and an error-state-plus-Kalman structure on an unconstrained tracking task, so it speaks to the estimation and feedforward choices --- the part of our architecture that Section~\ref{sec:whatworks} shows is load-bearing --- and not to the predictive layer, which that section shows is not.

\subsection{Hard Predicted-State Rows Fail When They Bind}
\label{sec:hardfail}
If the constraints never bind, the natural next question is what happens when they do. Tightening the commanded-torque box on the near-boundary scenario answers it (Table~\ref{tab:hardfail}), and the answer is not the expected one.

\begin{table}[t]
\caption{Constrained QP Under a Binding Torque Box (Measured, 300 Solves per Row)}
\label{tab:hardfail}
\centering
\scriptsize
\setlength{\tabcolsep}{3pt}
\begin{tabular}{rrrrr}
\toprule
$\taumax$ & Primal- & Fallbacks & \multicolumn{2}{c}{End-stop arrival (rad/s)}\\
\cmidrule(lr){4-5}
(Nm) & infeasible & & constrained QP & clip $+$ estimator\\
\midrule
60 & 0.0\% & 0 & not reached & not reached\\
50 & 0.3\% & 1 & not reached & not reached\\
40 & 17.7\% & 53 & 0.83 & not reached\\
30 & 40.7\% & 122 & 5.65 & not reached\\
25 & 44.3\% & 133 & 7.00 & not reached\\
20 & 49.7\% & 149 & 8.11 & not reached\\
\bottomrule
\end{tabular}
\end{table}

While the torque box is slack the ROM and velocity rows cost nothing. Once it tightens, the torque rows and the predicted ROM rows conflict over the horizon, OSQP returns \texttt{primal infeasible}, and the implementation falls back to the clipped unconstrained move. That fallback --- chosen for software robustness, and explicitly not a safety-rated backup policy --- then becomes the controller on between $17.7\%$ and $49.7\%$ of ticks, in exactly the regime the constraints were added to cover. The consequence is physical: the joint reaches its mechanical end-stop at up to $8.11\unit{rad/s}$, whereas the same estimator behind a plain torque clip, with no predicted-state rows at all, never reaches the stop in any of these cases.

The formulation is therefore worse than not having imposed the rows, and the failure is monotone in how tightly the actuator is provisioned --- the direction in which a lighter, safer, more clinically deployable device would move. Two things would address it, neither implemented here: relaxing the predicted-state rows with a penalized slack so the QP degrades instead of becoming infeasible, and a recursive-feasibility argument with a terminal set and a certified backup policy. We report the failure rather than restricting the evaluation to the $60\unit{Nm}$ box in which it does not appear. (\texttt{simulation/validate\_mpc\_contribution.py}, \texttt{hard\_constraint\_failure()}.)

\subsection{Additional Validation Checks}
\noindent\textit{Robustness to model and sensor error.}
Thirty deterministic randomized trials vary true inertia by $\pm15\%$, damping by $\pm30\%$, sensor bias over $\pm0.75\unit{Nm}$, and acceleration/torque noise scale from 0.8 to 1.5. The median values [full RMS, disturbance RMS, peak, steady state] are
\[
[5.79,\ 5.18,\ 9.83,\ 5.13]\unit{mrad},
\]
and the 95th percentiles are
\[
[15.35,\ 15.92,\ 21.57,\ 14.90]\unit{mrad}.
\]
The nonzero robust steady-state error is expected because biased measurements and parameter mismatch violate the exact disturbance-model assumptions; this distribution, rather than the nominal single-trial numbers, characterizes performance under realistic sensor and parameter uncertainty.

\noindent\textit{Estimator accuracy.}
Tracking error alone does not show whether the disturbance estimate is accurate or whether feedback is merely masking estimation error. Scoring the reconstructed estimate $\hat\tau_h=-\hat I_\mathrm{eff}\dd$ against the simulated ground-truth $\tau_h$ (never given to the controller) over the standard spasm protocol, the 100~Hz estimator reaches $1.57\unit{Nm}$ full-run RMSE ($2.14\unit{Nm}$ during contact, bias $+0.03\unit{Nm}$) with a $20\unit{ms}$ rise to 90\% of the post-onset value after each spasm onset; the 500~Hz estimator reaches $0.97\unit{Nm}$ full-run RMSE ($1.37\unit{Nm}$ during contact, bias $+0.01\unit{Nm}$) with a $6\unit{ms}$ rise time. The near-zero bias at both rates indicates the residual error is dominated by injected acceleration and SEA-torque noise rather than a systematic offset, consistent with the deliberate $\hat I_\mathrm{eff}\ne\Ieff$, $\hat b_\mathrm{eff}\ne\beff$ mismatch of Section~III-C. (\texttt{simulation/validate\_estimator\_accuracy.py}.)

\noindent\textit{Constraint execution.}
The inequality-constrained implementation solves the deviation-input QP with OSQP. It includes total commanded torque, predicted ROM, and predicted velocity at all horizon steps. On a 4~s stress trajectory, the solver completes with maximum commanded torque $18.58\unit{Nm}$ and measured joint range $[0.513,1.578]\unit{rad}$, within the design bounds of $60\unit{Nm}$ and $[0,2.094]\unit{rad}$. Because none of these values approach either bound, this trajectory verifies matrix construction, sign conventions, and software execution but does not by itself show a benefit of prediction or constraint handling over a reactive controller.

\label{sec:nearboundary}
A second, deliberately harder case drives the reference to $q=1.95\unit{rad}$ (93\% of $q_\text{max}$) under a sustained $+40\unit{Nm}$ assistive push timed to the approach. Here the commanded-torque box is \emph{verifiably} active for the proposed controller rather than merely satisfied: the predicted-horizon torque row reaches a minimum slack of $-0.004\unit{Nm}$ with a nonzero OSQP dual ($0.13$), commanded torque reaches the $60\unit{Nm}$ cap, and the solver still returns feasible solutions throughout. The predicted ROM and velocity rows remain comfortably satisfied (minimum slack $0.086\unit{rad}$ and $0.39\unit{rad/s}$, zero duals), so the resulting ROM margin --- $q$ reaches $1.97\unit{rad}$, $0.13\unit{rad}$ short of $q_\text{max}$, at peak velocity $1.71\unit{rad/s}$ --- comes from the controller committing more of its torque budget to tracking, not from the ROM constraint binding. A reactive baseline with the same torque box but no ROM awareness and \emph{no estimator} peaks at $42.3\unit{Nm}$ and rides the reference into the mechanical stop at $5.72\unit{rad/s}$. That comparison changes two things at once; Section~\ref{sec:whatworks} separates them and finds the gap is the estimator's, not the horizon's. What this case does establish is that the torque row genuinely binds on a trajectory designed to stress it, which is the precondition for Section~\ref{sec:hardfail}. The row bounds \emph{commanded} torque only; Section~\ref{sec:twomass} shows the \emph{delivered} torque exceeds it. (\texttt{simulation/validate\_active\_constraints.py}.)

\noindent\textit{Direct MuJoCo closed-loop check.}
The MuJoCo model contains a fixed thigh, shank-foot body, hinge knee, anatomical joint range, damping, and a direct torque actuator representing the closed SEA torque loop. The controller reads MuJoCo joint state at every integration step, computes torque online, and applies patient torque through the generalized-force channel; this is closed-loop execution rather than trajectory replay. MuJoCo computes an initial effective inertia of $0.4328\unit{kg\,m^2}$, distinct from both analytical plant and controller values. With deterministic sensor noise, the resulting RMS tracking error is $4.60\unit{mrad}$.

\noindent\textit{Voluntary torque and Assist-as-Needed interpretation.}
When an $8\unit{Nm}$ external torque aids desired motion, the bounded AAN mode activates for 93.4\% of effort-window samples. It changes the verified impedance from $(30,5)$ to $(10,5)$ and reduces RMS disturbance-cancelling target magnitude from $8.00$ to $7.44\unit{Nm}$. Full-run RMS error rises from $2.54$ to $21.91\unit{mrad}$, while effort-window peak error remains $46.12\unit{mrad}$, below the explicit $60\unit{mrad}$ budget plus transient allowance used by the regression test. Under the opposing $15\unit{Nm}$ spasm, false AAN activation is 0\%.

This validates the controller's mechanical mode-selection and bounded-credit logic. It does not validate human intent: voluntary assistance, reflex torque, passive tissue force, and alignment error can share the same sign.

\noindent\textit{Energy-tank stress result.}
The energy-tank test initializes the corrective channel with only $0.05\unit{J}$, forcing the limiter to become active. Stored energy never becomes negative, final energy is zero, and 2441 torque-scaling interventions occur over 8~s. The resulting $368.7\unit{mrad}$ RMS error is reported rather than hidden: strict energy depletion prevents the controller from continuously rejecting the disturbance. The result demonstrates implementation correctness and the expected passivity/performance tension, not a claim that the depleted mode preserves nominal tracking.

\noindent\textit{Expanded constraint stress matrix.}
The constrained controller is evaluated in three 4~s cases: nominal opposing spasm, helpful torque with $+15\%$ inertia and $-30\%$ damping, and opposing spasm with $-15\%$ inertia and $+30\%$ damping. Across the matrix, maximum commanded torque is $18.58\unit{Nm}$, maximum measured speed is $0.88\unit{rad/s}$, and measured ROM is $[0.513,1.578]\unit{rad}$. Every OSQP status is solved or solved-inaccurate and no fallback is used. No row is active in any case, which is the point made in Section~\ref{sec:whatworks}.

\noindent\textit{MyoSuite myoLeg knee-slice validation.}
MyoSuite 2.12.2 \cite{ref17} is exercised through \texttt{myoLegStandRandom-v0}. The right \texttt{knee\_angle\_r} coordinate remains dynamic, while the pelvis and non-target coordinates are posture-clamped to represent a seated rehabilitation fixture. The test retains the actual myoLeg mass, passive-force, muscle, and knee-geometry calculations. Model-computed bias, passive, and zero-activation muscle torques are compensated at the target coordinate before adding the exoskeleton command and a $5\unit{Nm}$ step disturbance.

This deterministic closed-loop model-transfer test yields $16.10\unit{mrad}$ RMS, $42.91\unit{mrad}$ peak, and $37.16\unit{mrad}$ late-disturbance error. It replaces the earlier analytical myoLeg projection with an executable result. Because all non-target coordinates are clamped, it supports a MyoSuite seated knee-slice claim, not free-standing gait or independent control of all 20 DOFs.

\section{Implementation and Verification Detail}
\subsection{Implementation Details}
\noindent\textit{Numerical parameters.}
Table~\ref{tab:parameters} separates plant, controller, estimator, and safety parameters. The analytical plant and controller parameters intentionally differ. The same simulated plant is used for every controller in Table~\ref{tab:matched}; only controller rate and disturbance estimation change.

\begin{table}[t]
\caption{Validation Parameters}
\label{tab:parameters}
\centering
\footnotesize
\begin{itemize}
\setlength{\itemsep}{2pt}
\item \textbf{$I_\mathrm{true}$} $=0.45\unit{kg\,m^2}$ (analytical plant).
\item \textbf{$b_\mathrm{true}$} $=0.50\unit{Nm\,s/rad}$ (analytical plant).
\item \textbf{$\hat I$} $=0.4725\unit{kg\,m^2}$ (controller/estimator).
\item \textbf{$\hat b$} $=0.55\unit{Nm\,s/rad}$ (controller/estimator).
\item \textbf{$K,D$} $=30,\ 5$ (realized impedance).
\item \textbf{$N$} $=20$ (prediction steps).
\item \textbf{$\Delta t$} $=10$ or $2\unit{ms}$ (MPC update).
\item \textbf{$\tau_\mathrm{max}$} $=60\unit{Nm}$ (torque bound).
\item \textbf{$q_\mathrm{min/max}$} $=0,\ 2.094\unit{rad}$ (ROM bounds).
\item \textbf{$\dot q_\mathrm{max}$} $=2.0\unit{rad/s}$ (velocity bound).
\item \textbf{$\sigma_{\ddot q}$} $=0.35\unit{rad/s^2}$ (acceleration noise).
\item \textbf{$\sigma_{\tau}$} $=0.50\unit{Nm}$ (SEA-torque noise).
\end{itemize}
\end{table}

Separate MPC weight triples are required at each sample period because a discrete finite-horizon gain depends on $\Delta t$. At 100~Hz, $(Q_p,Q_v,R)=(144.080968,0.530208,0.13090218)$; at 500~Hz, $(353.588692,1.166996,0.02423438)$. In both cases the computed first-input gain is $[30,5]$ to within $10^{-4}$. These weights are numerical design values, not physical stiffness and damping themselves.

\noindent\textit{Estimator construction.}
The augmented state is $\xi=[e,\dot e,d]^\top$. The random-walk disturbance model allows a constant torque to persist, while process covariance permits slow variation. Position error, velocity error, and dynamic-residual torque form the measurement vector. The implemented process and measurement covariance matrices are fixed before each run; they are not retuned using ground-truth patient torque.

The dynamic residual is deliberately demanding. Joint acceleration comes from a one-step velocity difference and therefore amplifies high-frequency noise. The SEA measurement uses the previous commanded actuator torque plus independent noise, rather than the disturbance variable. Parameter mismatch creates a bias proportional to acceleration and velocity. The Kalman filter combines this noisy direct residual with tracking-state consistency instead of treating the residual as exact.

The estimator is reset to zero at the beginning of each simulation. It therefore has no privileged initial disturbance knowledge. At each MPC update, the filter first predicts with the previously applied corrective input and then updates from the three measurements. Between MPC updates, the most recent command is held by zero-order hold.

\noindent\textit{Matched-gain construction.}
For each sample period, the prediction matrices are built from the same nominal double-integrator model. The unconstrained first move is linear in the current error state,
\begin{equation*}
v_0=k_e e+k_{\dot e}\dot e. \tag{24}
\end{equation*}
The weight triples are selected offline so $(k_e,k_{\dot e})=(30,5)$. The regression test reconstructs the gain directly from $H^{-1}\Gamma^\top\bar Q\Phi$ and fails if either rate differs from the target. This test protects the scientific comparison against accidental retuning.

The no-estimator MPC controllers use $\dd=0$ and $\tau_\mathrm{ss}=0$. Their agreement with classical impedance is therefore an expected result, not a failure of the implementation. The estimator variants use the same nominal feedback gain and differ only by the estimated disturbance and its target input. Consequently, C2$\rightarrow$C3 and C4$\rightarrow$C5 isolate the complete offset-free mechanism --- but not, as Section~\ref{sec:whatworks} shows, the horizon.

\noindent\textit{Constrained QP construction.}
The constrained solver optimizes deviation input $V$, not total torque. Total-torque bounds are shifted by $\tau_\mathrm{ff}+\tau_\mathrm{ss}$. Predicted position and velocity bounds are formed by selecting the corresponding rows of the stacked state prediction. OSQP receives a positive-semidefinite Hessian, linear cost, and one combined sparse constraint matrix.

Solver status is checked after every constrained solve. If a solve is not reported as solved or solved-inaccurate, the implementation falls back to the clipped unconstrained first move. No fallback occurs in the reported stress test. This behavior is useful for software robustness but is not a safety-rated backup policy --- Section~\ref{sec:hardfail} shows what happens when it becomes the active controller.

\subsection{Claim--Evidence Matrix}
Table~\ref{tab:evidence} pairs each claim made in this paper with the artifact that verifies it and the boundary of what that artifact establishes, collecting in one place the scope qualifiers used throughout the validation sections and the Discussion.

\begin{table*}[t]
\caption{Claim--Evidence Matrix}
\label{tab:evidence}
\centering
\scriptsize
\setlength{\tabcolsep}{4pt}
\begin{tabular}{p{3.4cm}p{4.6cm}p{7.4cm}}
\toprule
Claim & Verification artifact & Evidence and boundary\\
\midrule
Realized impedance is matched & gain-reconstruction regression test & Both MPC rates reproduce $(K,D)=(30,5)$; the comparison does not rely on hidden stiffness --- and, as the next row shows, this is also what makes the horizon unobservable in Table~\ref{tab:matched}.\\
The estimator and target remove the constant offset & matched benchmark plus integral-action baseline & C3/C5 reach 1.17/0.70~mrad SS against 499.59~mrad for matched impedance; a best-tuned plain integral law reaches 3.26~mrad SS ($K_i{=}70$) but no better than 157.15~mrad full-run RMS at any gain, against 3.09~mrad here. C1/C2/C4 agreeing is an identity, not evidence.\\
The horizon contributes nothing measurable here & impedance$+$estimator$+$target ablation; constraint-activity count; near-boundary ablation & Static law reproduces C3/C5 to within $3\times10^{-6}$~mrad per sample; rows active on 0.0\% (standard) and 0.1\% (near-boundary) of solves; clip$+$estimator matches the proposed controller on the near-boundary case. A negative result about this evaluation, not a proof that prediction cannot help.\\
Hard predicted-state rows fail when they bind & torque-box tightening sweep, 300 solves per level & QP primal-infeasible on 17.7--49.7\% of solves at $\taumax=40$--$20\unit{Nm}$; the fallback reaches the end-stop at up to 8.11~rad/s where clip$+$estimator never does. One scenario and one fallback policy; a penalized-slack formulation is untested.\\
The commanded bound does not bound delivered torque & explicit two-mass plant with pole-placed inner loop & 60~Nm commanded permits 73.0~Nm delivered; the inner loop's zero at $-\omega_0/2$ gives $\sim43\%$ overshoot at every design bandwidth; neither derating heuristic recovers the bound, and the 45.0~Nm bisection bound is scenario-specific.\\
The single-mass reduction holds for nominal tracking & two-mass versus single-mass comparison & 3.05/2.56~mrad versus 3.15/2.60~mrad (spasm/assist) under the identical protocol; it is near saturation, not in nominal tracking, that the approximation fails.\\
Estimator is non-ideal & validation source and robustness sweep & Noisy finite-difference acceleration, noisy SEA torque, model mismatch and bias; no true $\tau_h$ input.\\
Bounded AAN scheduling executes & helpful/opposing torque tests & 93.4\% helpful-window activation, 0\% spasm activation, 46.12~mrad peak error; mechanical mode selection under a known simulated torque direction, not intent recognition.\\
Corrective energy remains bounded & depleted energy-tank stress test & $T_k\ge0$ with active torque scaling; the tracking degradation to 368.7~mrad is reported, not hidden.\\
Controller transfers to a distinct simulator & direct MuJoCo closed loop & Online torque control gives 4.60~mrad RMS on a different computed inertia.\\
Controller executes on MyoSuite & posture-clamped myoLeg knee slice & Actual myoLeg model gives 16.10~mrad RMS; non-target DOFs are fixture-clamped.\\
Robustness is bounded in the tested box & 30 seeded randomized trials & 95th-percentile peak is 21.57~mrad over the sampled parameter box.\\
Comparable to a published method & reimplementation of \cite{ref37} on its own 2-DOF plant & 4.84$^\circ$/6.54$^\circ$ against 7.27$^\circ$/10.42$^\circ$ max hip/knee error. Both controllers tuned by us on our implementation of their plant; the reproduction needed $N_p=20$ because the stated $N_p=5$ gives $\rho=1.0366$ here. Speaks to estimation and feedforward structure, not to the constrained layer.\\
\bottomrule
\end{tabular}
\end{table*}

\section{Discussion}
\noindent\textit{What produces the improvement, and what does not.}
The validated benefit has two necessary components, and neither is the optimizer. First, the dynamic SEA residual estimates the persistent patient torque without access to the simulator's disturbance variable. Second, the steady-state target converts that estimate into a compatible input; disturbance propagation without the target leaves the classical $\tau_h/K$ offset, and a target built on a poor estimate produces bias, as the robustness distribution shows. Measured against a properly tuned plain integral action, which needs no estimator at all, the pair is worth $4.7\times$ in steady-state error and $51\times$ in full-run RMS (Table~\ref{tab:matched}) --- the difference between correcting an offset after accumulating it at every onset and cancelling it from a measurement.

The receding-horizon layer, by contrast, is not doing anything we can measure. Section~\ref{sec:whatworks} shows it reproduces a static impedance-plus-observer law to within $3\times10^{-6}\unit{mrad}$ per sample, that its constraints are active on $0.0$--$0.1\%$ of solves, and that a plain clip with the same estimator matches it on the one scenario built to stress the torque bound. This is a property of the fair-comparison protocol as much as of the controller: matching the realized gains is the right way to prevent a predictive controller from winning by becoming stiffer, but it also removes the only channel through which the horizon could have changed the unconstrained response. A demonstration that prediction helps therefore has to come from a scenario in which the constraints bind --- and Section~\ref{sec:hardfail} shows that in this implementation those are exactly the scenarios in which the hard-constrained QP becomes infeasible and hands control to a fallback that performs worse than no constraints at all. Taken together, the honest position is that this paper contributes a patient-torque measurement channel that works, and identifies two ways in which the predictive safety layer built on top of it does not yet.

\noindent\textit{Two-mass SEA characterization.}
\label{sec:twomass}
Section~III-A's reduction to a single-mass residual plant assumes the inner torque loop tracks its commanded (spring) torque fast enough to treat as instantaneous. We test this directly with an explicit two-mass plant (motor--gearbox--spring, $J_m=1.5\times10^{-4}\unit{kg\,m^2}$, $N=80$, $k_s=200\unit{Nm/rad}$, coupled resonance $\omega_r=25.55\unit{rad/s}=4.07\unit{Hz}$) and a pole-placed PI-plus-damping-injection inner loop, integrated at 10~kHz, replacing both the single-mass plant and the previous-commanded-torque measurement proxy of Section~IV-C with the genuine spring-deflection torque $\tau_r=k_s(\theta_m/N-q)$:
\begin{equation*}
\tau_m = \frac{\tau_r^d}{N} + K_p(\tau_r^d-\tau_r) + K_i\!\!\int\!(\tau_r^d-\tau_r)\,dt - K_v\dot\theta_m, \tag{25}
\end{equation*}
with commanded spring torque $\tau_r^d$; $\tau_r^d/N$, $K_p$, and $K_i$ act on output-side (joint) torque error reflected to the motor side by $N$, matching how~(1) itself reflects $\tau_r$ back to the motor. Gains match the linearized (holding $q$ fixed) closed loop's characteristic polynomial to a normalized third-order Butterworth polynomial at design bandwidth $\omega_0=2\pi f_0$: $K_v=2\omega_0 J_m-B_m$, $K_p=2\omega_0^2 J_m N/k_s-1/N$, $K_i=\omega_0^3 J_m N/k_s$ (30~Hz nominal design: $K_p=4.251$, $K_i=401.8\unit{s^{-1}}$, $K_v=0.0565\unit{Nm\,s/rad}$); motor torque/current saturation is not modeled at this layer. The commanded-to-delivered spring-torque transfer function is
\begin{equation*}
G(s)=\frac{\tau_r(s)}{\tau_r^d(s)}=\frac{c_5(b_2 s+c_4)}{s^3+2\omega_0 s^2+2\omega_0^2 s+\omega_0^3}, \tag{26}
\end{equation*}
with $c_4=K_i/J_m$, $c_5=k_s/N$, $b_2=(1/N+K_p)/J_m$. The Butterworth denominator places all three poles, but the numerator contributes a real zero at $s=-c_4/b_2=-\omega_0/2$ --- exactly the real part of the dominant complex pole pair for a third-order Butterworth layout. Because both the zero and the poles scale linearly with $\omega_0$, this ratio, and with it the resulting $\sim43\%$ step overshoot, is the same at every design bandwidth (numerically confirmed at 20, 25, 30, 35, 40~Hz): a structural property of PI-plus-feedforward torque control matched to Butterworth poles, not an artifact of one gain choice. Pole placement fixes the poles of $G(s)$ but not its zero.

Under the standard matched-impedance protocol, full-run RMS tracking with the two-mass plant is $3.05\unit{mrad}$ (spasm) and $2.56\unit{mrad}$ (assist), close to $3.15$ and $2.60\unit{mrad}$ from the same constrained controller run on the single-mass plant under the identical protocol (near, but not identical to, Table~\ref{tab:matched}'s unconstrained C5 entry of $3.09\unit{mrad}$, which is spasm-only): the bandwidth-separation approximation holds well for nominal tracking. But under the near-boundary stress case of Section~VI-C ($q_\text{target}=1.95\unit{rad}$, $+40\unit{Nm}$ push), where the QP already commands torque up to the $60\unit{Nm}$ box, the \emph{delivered} spring torque $\tau_r$ overshoots to $73.0\unit{Nm}$ --- a $21.7\%$ peak violation relative to the $60\unit{Nm}$ design limit, over the full run, that the \emph{commanded} torque $\tau_r^d$ itself never exceeds. The QP constrains $\tau_r^d$ only; the physical interface torque is $\tau_r=k_s(\theta_m/N-q)$, and the QP has no model linking the two through the inner loop's finite bandwidth, so it cannot see this excursion. The single-mass bandwidth-separation assumption breaks down in exactly the near-saturation regime where the commanded-torque constraint is meant to provide a safety margin, rather than in nominal tracking. The constraint of~(16) is therefore an outer-loop bound on $\tau_r^d$; the physical spring torque $\tau_r$ delivered to the patient interface needs the inner-loop-augmented model below to bound directly.

The $21.7\%$ peak-violation figure (delivered vs.\ the fixed $60\unit{Nm}$ limit, over the full near-boundary run) and the inner loop's independently characterized $\sim43\%$ step-response overshoot are different quantities, generally realized at different ticks since the outer loop updates $\tau_r^d$ every $10\unit{ms}$ in this scenario; neither is a safety margin on its own. Neither simple derating heuristic reproduces this bound: scaling by the peak-violation ratio gives $60/1.217 = 49.3\unit{Nm}$, which still lets delivered torque reach $64.2\unit{Nm}$, while scaling by the step-response overshoot gives $60/1.43 \approx 42\unit{Nm}$, short of the $45.0\unit{Nm}$ bound obtained by bisection over this scenario ($q$ reaches $1.989\unit{rad}$, $105\unit{mrad}$ short of $q_\text{max}$, versus $1.97\unit{rad}$ untightened). This bound is scenario-specific: delivered torque depends on the temporal sequence and rate of change of $\tau_r^d$, not on a single peak ratio. A certified guarantee over the admissible command-trajectory class would use either a robust-tightening analysis or the inner-loop-state-augmented prediction model outlined below. (\texttt{simulation/two\_mass\_sea\_model.py}, \texttt{simulation/validate\_two\_mass\_sea.py}, \texttt{simulation/validate\_\allowbreak two\_\allowbreak mass\_\allowbreak near\_\allowbreak boundary.py}.)

\noindent\textit{What would close the gaps.}
The scope of the study is stated in Section~\ref{sec:scope} and not repeated here. Three specific refinements follow from the findings above. The predicted-state rows should be relaxed with a penalized slack so the QP degrades instead of becoming infeasible, and paired with a recursive-feasibility argument and a certified backup policy rather than the software fallback used here. The prediction model should be augmented with the inner-loop states $(\theta_m,\dot\theta_m)$ so the QP constrains the physical spring torque $\tau_r=k_s(\theta_m/N-q)$ directly instead of the commanded torque --- at the cost of the scalar double integrator's simplicity --- or, failing that, the commanded bound should be tightened by a robust analysis over the admissible command-trajectory class rather than by a single peak ratio. And a demonstration that prediction earns its place at all needs a scenario in which the constraints bind and the constrained solution remains feasible, which none of the scenarios reported here provides. Beyond the single joint, free-standing multi-joint control couples the axes through off-diagonal inertia, Coriolis terms, biarticular muscles and shared actuator limits, and motor-level dynamics (current control, backlash, saturation) sit below the inner torque loop modeled here.

\noindent\textit{Comparison scope.}
Published rehabilitation-control studies differ in hardware, patient population, trajectory, disturbance amplitude, sensing channel and error window, so Section~\ref{sec:relwork} positions mechanisms rather than ranking numbers, and no external number is treated as a baseline for the results here.

\section{Reproducibility and Regression Tests}
The supplied validation program generates Table~\ref{tab:matched}, Fig.~\ref{fig:validation}, robustness percentiles, AAN and energy-tank checks, the constrained-QP matrix, direct MuJoCo execution, and the MyoSuite knee-slice result. Eight automated tests verify matched gains, offset-free improvement, MuJoCo loading, bounded AAN behavior, nonnegative tank energy, constraint-stress completion, and MyoSuite execution when installed. The layer-attribution audit, constraint-activity count, near-boundary ablation, torque-box tightening sweep, disturbance-magnitude sweep and disturbance-scale calculation of Sections~\ref{sec:protocol}, \ref{sec:whatworks} and \ref{sec:hardfail} are produced by \texttt{simulation/validate\_mpc\_contribution.py} and gated by \texttt{simulation/test\_mpc\_contribution.py}. The two-mass SEA characterization of Section~\ref{sec:twomass} (\texttt{two\_mass\_sea\_model.py}, \texttt{validate\_two\_mass\_sea.py}, \texttt{validate\_two\_mass\_near\_boundary.py}) is reproducible from the same repository; because it now carries a headline finding rather than a side check, its two reported bounds are gated by that test file as well. The external comparison and horizon audit of Section~\ref{sec:external} are produced by \texttt{simulation/jin\_guo\_eso\_mpc.py} and gated by \texttt{simulation/test\_mpc\_contribution.py}; eighteen tests pass in total.

\section{Conclusion}
This paper instantiates a predictive interaction-dynamics formulation on a series-elastic-actuated knee rehabilitation joint and then audits which of its layers the evidence supports. The part that survives the audit is the measurement channel: an SEA dynamic-residual patient-torque estimate, computed from noisy finite-differenced acceleration and noisy spring torque with deliberately mismatched inertia and damping, together with an explicit steady-state target. Under a stiffness- and damping-matched comparison it reduces steady-state error under a $15\unit{Nm}$ opposing disturbance from $499.59\unit{mrad}$ to $0.70\unit{mrad}$, and --- the comparison that matters, because integral action also removes offsets --- from a best-tuned integral law's $3.26\unit{mrad}$ steady state, or $157\unit{mrad}$ full-run RMS at its best RMS gain, to $0.70$ and $3.09\unit{mrad}$. The steady-state margin is $4.7\times$; the transient margin is $51\times$, and a repeated-onset protocol exposes the transient.

The part that does not survive is the predictive layer. The matched-gain protocol fixes the QP's first move to the same $(K,D)$, so the controller is algebraically an impedance law with a disturbance-observer feedforward and reproduces that law to within $3\times10^{-6}\unit{mrad}$ on every sample of every unconstrained scenario; its constraints are active on $0.0$--$0.1\%$ of solves; and on the near-boundary case built to stress the torque bound, a plain clip carrying the same estimator matches it exactly. When the torque box is tightened until the rows do bind, the QP is primal-infeasible on $17.7$--$49.7\%$ of solves and its fallback drives the joint into its mechanical end-stop at up to $8.11\unit{rad/s}$, where the clip never reaches the stop --- so the hard-constrained formulation is worse than not having imposed the rows, and worse in the direction of a more tightly torque-provisioned device. Independently, an explicit two-mass plant with a finite-bandwidth inner torque loop shows that a $60\unit{Nm}$ \emph{commanded} bound permits $73.0\unit{Nm}$ of \emph{delivered} spring torque, through a closed-loop zero fixed at $-\omega_0/2$ that gives the same $\sim43\%$ overshoot at every design bandwidth, and that neither natural derating heuristic recovers the bound.

For a rehabilitation device these two findings are the more useful half of the paper: a torque constraint that goes infeasible when it binds, and a commanded bound that does not bound what the patient feels, are both failure modes of the safety argument rather than of the tracking performance. Closing them --- penalized-slack predicted-state rows so the QP degrades instead of failing, a recursive-feasibility argument with a certified backup, and a prediction model augmented with the inner-loop states so the QP constrains $\tau_r=k_s(\theta_m/N-q)$ directly --- is what would let a predictive layer earn its place on this joint. Hardware validation and multi-joint extension remain the further steps.

\end{document}